\documentclass[twocolumn,showpacs,preprintnumbers,amsmath,amssymb,prl,
superscriptaddress]{revtex4}

\usepackage{graphicx}
\usepackage{dcolumn}
\usepackage{bm}

\input def.sty

\begin{document}


\title{Precision Measurements of the Nucleon Strange Form Factors 
at $Q^2\sim 0.1\gevc$}

%
%

\author{A.~Acha}
\affiliation{ \mbox{Florida International University}, Miami, Florida 33199, USA }

\author{K.~A.~Aniol}
 \affiliation{ \mbox{California State University, Los Angeles}, Los Angeles, California 90032, USA }

\author{D.~S.~Armstrong}
\affiliation{College of William and Mary, Williamsburg, Virginia 23187, USA}

\author{J.~Arrington}
\affiliation{Argonne National Laboratory, Argonne, Illinois 60439, USA}

\author{T.~Averett}
\affiliation{College of William and Mary, Williamsburg, Virginia 23187, USA}

\author{S.~L.~Bailey}
\affiliation{College of William and Mary, Williamsburg, Virginia 23187, USA}

\author{J.~Barber}
\affiliation{University of Massachusetts Amherst, Amherst, Massachusetts 01003, USA}

\author{A.~Beck}
\affiliation{Massachusetts Institute of Technology, Cambridge, Massachusetts 02139, USA} 

\author{H.~Benaoum}
\affiliation{Syracuse University, Syracuse, New York 13244, USA} 

\author{J.~Benesch}
\affiliation{Thomas Jefferson National Accelerator Facility, Newport News, Virginia 23606, USA} 

% ...

\collaboration{The HAPPEX Collaboration}
\noaffiliation
%
%

\date{September 4, 2006}

\begin{abstract}
We report  
new measurements of the parity-violating asymmetry $\APV$ in 
elastic scattering of 3~GeV electrons off hydrogen and $\he$ targets with
$\langle\theta_{lab}\rangle\approx 6.0^{\circ}$. The
$^4$He result is 
$\APV =  (+6.40 \pm 0.23 \,\,\mbox{(stat)} \,\pm 0.12 \,\,\mbox{(syst)})
\times 10^{-6}$. The hydrogen result is
$\APV =  (-1.58 \pm 0.12 \,\,\mbox{(stat)} \,\pm 0.04 \,\,\mbox{(syst)})
\times 10^{-6}$. 
These results significantly improve constraints on the electric and magnetic
strange form factors $\ges$ and $\gms$.
We extract $G^{s}_{E} = 0.002 \, \pm \, 0.014 \, \pm 0.007$
at $\langle\qsq\rangle = 0.077\gevc$, 
and $G^{s}_{E} + 0.09 \, G^{s}_{M} = 0.007 \, \pm \, 0.011 \, \pm 0.006$
at $\langle\qsq\rangle = 0.109\gevc$, providing new limits on the 
role of strange quarks in the nucleon charge and magnetization distributions. 

\end{abstract}

\maketitle


Over the past several decades, high-energy lepton-nucleon scattering has 
revealed the rich structure of the nucleon 
over a wide range of length 
scales. In recent years, increasingly sensitive measurements of elastic
electron-nucleon scattering, mediated by photon exchange and Z$^0$
exchange, have enabled the measurement of the electromagnetic and neutral
weak form factors. These functions of the 4-momentum transfer $Q^2$
characterize nucleon charge and magnetization distributions.

In particular, the neutral weak form factor measurements provide a way
to probe dynamics of the ``sea'' of virtual light (up, down and strange)
quark-antiquark pairs that 
surrounds each valence quark in the nucleon. Since the Z$^0$ boson couples
to various 
quarks with different relative strengths compared to the photon, 
a combined analysis of proton and neutron electromagnetic
form factor and proton neutral weak form factor measurements, along with the
assumption of charge symmetry, allows the determination of the strange
electric and magnetic form factors  $G_E^s$ and $G_M^s$~\cite{aneesh,bob}. 

The established experimental technique to measure
the electron-nucleon weak neutral current amplitude is parity-violating
electron scattering~\cite{prescott, musolf94}. Longitudinally-polarized 
electron-scattering off unpolarized targets can access a 
parity-violating asymmetry
$\APV\equiv{(\sigma_R-\sigma_L)}/{(\sigma_R+\sigma_L)}$, where
$\sigma_{R(L)}$ is the cross section for incident right(left)-handed
electrons.  Arising from the interference of the weak and
electromagnetic amplitudes, $\APV$ increases with $\qsq$~\cite{zeld}. 

Four experimental programs have been designed to access the
$\qsq$ range of 0.1 to 1$\gevc$, where the $\APV$ expectations 
range from one to tens of parts per million (ppm).  The published
measurements~\cite{sample,happex, A41, A42, happexhe,happexh,G0} are
mutually consistent.
An intriguing pattern in the low-$\qsq$ behavior seen in~\cite{A42,G0} 
has marginal statistical significance. 

In this paper, we significantly improve our two previous
measurements~\cite{happexhe,happexh} of $\APV$ in 
elastic electron scattering from $\h$ and $\he$ nuclei.
Since $\APV$ for $\h$ is sensitive to a linear combination of $\ges$ and $\gms$
while that for $\he$ is sensitive only to $\ges$, a simultaneous
analysis of both measurements results in the most precise
determination to date of $\ges$ and $\gms$ at $\qsq\sim0.1\gevc$.

The measurements were carried out in Hall A at the Thomas Jefferson National 
Accelerator Facility (JLab). 
As described in detail in two previous publications~\cite{happexhe,happexh},
a 35 to 55~$\mu$A continuous-wave beam of $\sim$3~GeV longitudinally polarized 
electrons was incident on 20~cm long cryogenic targets. 
Elastically scattered electrons were focused into background-free regions 
by a symmetric pair of high-resolution spectrometer systems. 
The scattered flux was intercepted by identical detector segments in each
arm (two for $\h$, one for $\he$), resulting in Cherenkov light
collected by photomultiplier tubes (PMTs).

The helicity of the electron beam, generated by photoemission off a GaAs
wafer, is determined by the handedness of the incident laser light's
circular polarization.  
This was selected pseudorandomly at 15~Hz and toggled to the opposing
helicity after 33.3~ms, with each of these equal periods of constant helicity 
referred to as a ``window.''  
PMT and beam monitor responses for two consecutive windows of opposite 
helicity were integrated, digitized, and grouped as a ``pair'' for
asymmetry analysis.

The beam monitors, target, detector components, electronics and
accelerator tune were optimized such that
the fluctuation 
in the PMT response 
over a pair was dominated by counting statistics of the scattered flux 
for rates up to 100 MHz.
This facilitated $\APV$ measurements with statistical uncertainty
as small as 100 parts per billion (ppb) in a reasonable length of time.
To keep spurious beam-induced asymmetries under control at this level, 
the laser optics 
leading to the photocathode were carefully designed and monitored.
Indeed, averaged over the entire period of data collection with the hydrogen target,
the achieved level of control surpassed all previous benchmarks, as summarized
in Table~\ref{tab:beam}. 
\begin{table}
\begin{tabular}{|l|rl|rl|} \hline
&  \multicolumn{2}{c|}{Helium} & \multicolumn{2}{c|}{Hydrogen} \\ \hline \hline
A$_{\rm{Intensity}}$             &   \, -0.377& ppm & \,\,\,0.406 &ppm   \\
A$_{\rm{Energy}}$ &     3 &ppb   &   0.2 &ppb     \\
$\Delta x$          &     -0.2 &nm  & 0.5 &nm       \\
$\Delta x^{\prime}$ &     4.4& nrad  &  -0.2 &nrad  \\
$\Delta y$          &    -26 &nm    &    1.7 &nm	  \\
$\Delta y^{\prime}$ &    -4.4 &nrad &    0.2 &nrad  \\  \hline
\end{tabular}
\caption{Average beam asymmetries under polarization reversal
in intensity and energy and differences in horizontal and vertical position
($\Delta x$, $\Delta y$) and angle ($\Delta x^{\prime}$, $\Delta y^{\prime}$) . }
\label{tab:beam}
\end{table}

The data collection took place over 55 days ($\he$) and 36 days ($\h$).
A half-wave ($\lambda$/2) plate was periodically inserted into the 
laser optical path which passively reversed the
sign of the electron beam polarization. 
With roughly equal statistics in each state, many systematic effects
were suppressed.
There were 121 ($\he$) and 41 ($\h$) such reversals.
The data set between two successive 
$\lambda/2$ reversals is referred to as a ``slug.''

Loose requirements were imposed on beam quality to
remove periods of instability, leaving
about 95\%\ of the data sample for further
analysis. No helicity-dependent cuts were applied.
The final data sample 
consisted of $35.0\times10^6$ ($\he$) and $26.4\times10^6$ ($\h$) pairs.
The right-left helicity asymmetry in the integrated detector response,
normalized to the beam intensity,
was computed for each pair to form the raw asymmetry
$\ARAW$. The dependence of $\ARAW$ on  
fluctuations in the five correlated beam parameter differences $\Delta x_i$ 
is quantified as $\ABEAM=\sum c_i\Delta x_i$, 
where the coefficients $c_i$ quantify the $\ARAW$ beam parameter sensitivity.
The electroweak physics of the
signal and backgrounds is contained in $\ACORR=\ARAW-\ABEAM$. 

The $\ACORR$ window-pair distributions for the two 
complete data samples were perfectly Gaussian over more than 4 
orders of magnitude with RMS widths of 1130 ppm ($\he$) and 540 ppm ($\h$); 
the dominant source of noise in the PMT response was counting statistics.  
To further test that 
the data behaved statistically and the errors were being accurately calculated,
$\ACORR$ averages and statistical errors for typical one hour runs,
consisting of about 50k pairs each, were studied. 
Each set of roughly 400 average $\ACORR$ values, normalized
by the corresponding statistical errors, populated a Gaussian 
distribution of unit variance as expected.

Systematic effects in $\ABEAM$ estimations
were studied. When averaged over all detector segments, 
the coefficients $c_i$ were much smaller 
than those for individual detector segments due to the
symmetric geometry of the apparatus. Limits on systematic uncertainties in the
$c_i$'s
in the range of 10 to 30\%\ were set by inspecting residual correlations
of $\ACORR$'s of individual detector segments with helicity-correlated beam 
asymmetries.

Another important validation was to use two 
independent methods to calculate $c_i$. The first relied on linear regression
of the observed response of the detector PMTs to  
intrinsic beam fluctuations. 
The other used calibration data in which the beam was modulated,
by amounts large compared to intrinsic beam fluctuations,
using steering magnets and an accelerating cavity.
Differences in the two $\ABEAM$ calculations were always
much smaller than corresponding $\ACORR$ statistical errors.

Final $\ACORR$ results were calculated using the beam modulation technique 
and are summarized in Table~\ref{tab:araw}.
Due to the excellent control of beam parameter
differences $\Delta x_i$ summarized in Table~\ref{tab:beam},
$\ACORR-\ARAW$ values 
are of the order of, or much smaller than,
the corresponding statistical errors.
Under $\lambda/2$ reversal, 
the absolute values of $\ACORR$ are consistent within statistical errors. 
The reduced $\chi^2$ for $\ACORR$ ``slug'' averages
is close to one in every case, indicating that any residual 
beam-related systematic effects were small and randomized over the time 
period of $\lambda/2$ reversals (typically 5 to 10 hours). 
The final $\ACORR$ results
are $\ACORR^{\rm{He}}=+5.25\pm 0.19\rm{(stat)}\pm 0.05\rm{(syst)}$ ppm 
and  $\ACORR^{\rm{H}}=-1.42\pm 0.11\rm{(stat)}\pm 0.02\rm{(syst)}$ ppm.

\begin{table}
\begin{tabular}{|r|c|c|c|c|c|c|}\hline
& \multicolumn{2}{c|}{$\lambda/2$ {\bf OUT}} & \multicolumn{2}{c|}{$\lambda/2$
{\bf IN}} & \multicolumn{2}{c|}{\bf BOTH} \\ \hline \hline
$^4$He & \multicolumn{2}{c|}{(DOF =  59)} & \multicolumn{2}{c|}{(DOF =  60)} &
\multicolumn{2}{c|}{(DOF = 120)} \\
 & Asym & r$\chi^2$ & Asym & r$\chi^2$ & Asym & r$\chi^2$ \\
\hline 
 $\ARAW$  & 4.80$\pm$0.27 & 0.75 & -5.41$\pm$0.27 & 1.12 & 5.10$\pm$0.19 
& 0.95 \\
 $\ACORR$  & 5.12$\pm$0.27 & 0.78 & -5.38$\pm$0.27 & 1.07 & 5.25$\pm$0.19 
& 0.92 \\ \hline \hline
$^1$H & \multicolumn{2}{c|}{(DOF =  20)} & \multicolumn{2}{c|}{(DOF =  19)} &
\multicolumn{2}{c|}{(DOF = 40)} \\
\hline 
 $\ARAW$  & -1.40$\pm$0.15 & 0.73 & 1.42$\pm$0.15 & 1.04 & -1.41$\pm$0.11 
& 0.86 \\ 
 $\ACORR$  & -1.41$\pm$0.15 & 0.81 & 1.43$\pm$0.15 & 1.02 & -1.42$\pm$0.11 
& 0.89 \\
\hline
\end{tabular}
\caption{Raw and corrected asymmetries (in ppm) and reduced 
``slug'' $\chi^2$ 
(r$\chi^2$), broken up by $\lambda/2$ reversals.  The differences between
$\ARAW$ and $\ACORR$ result from corrections for energy, position, and 
angle differences which are summarized in Table~\ref{tab:beam}.}
\label{tab:araw}
\end{table}

\begin{figure}
\includegraphics[width=1.0\columnwidth]{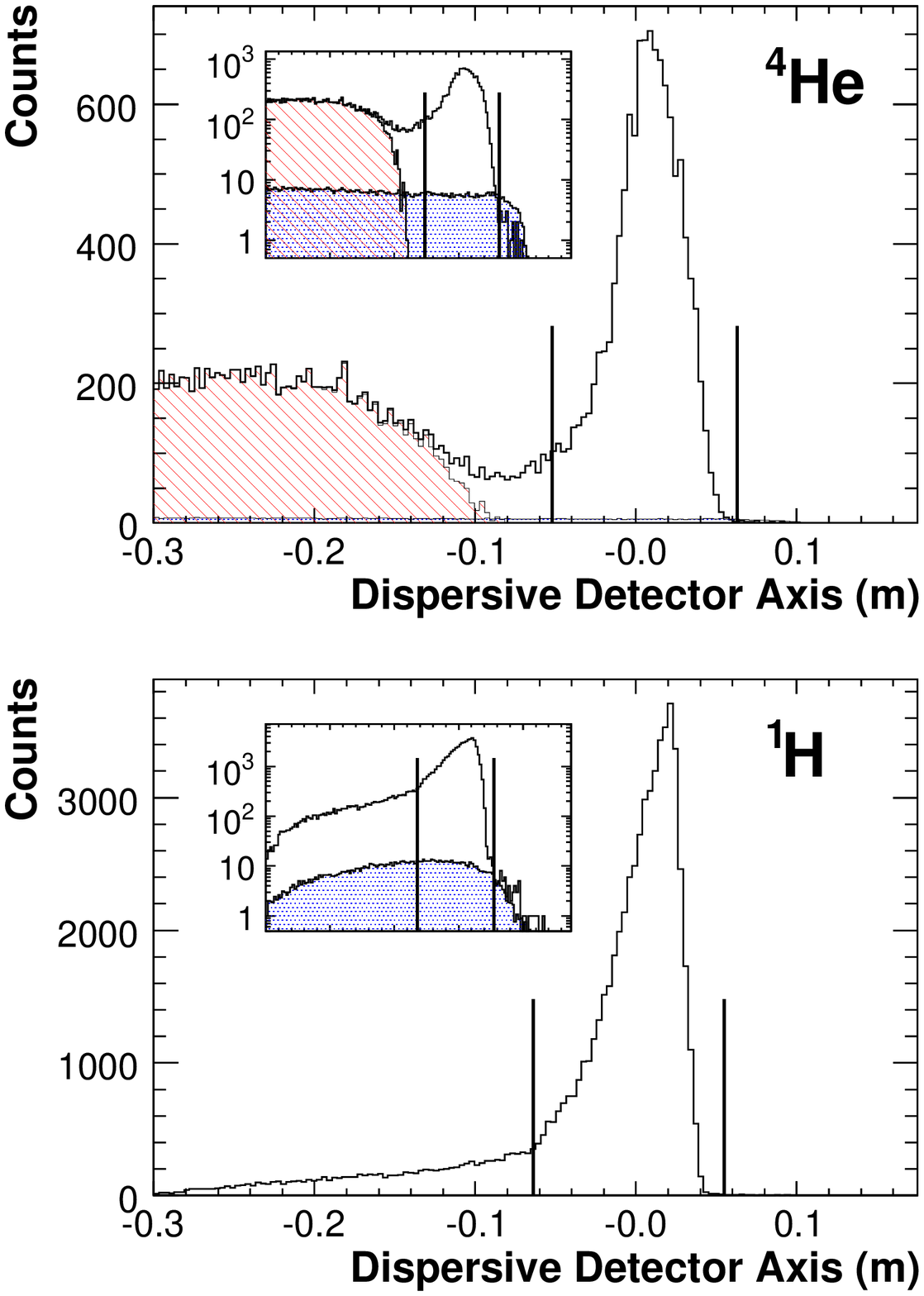}
\caption{Single-particle spectra obtained in dedicated low-current runs.
The insets show the same spectra on a logarithmic scale.  
The vertical lines delineate the extent of the detectors. 
Inelastic scattering from $\he$ is entirely contained in the hatched area.
The shaded regions, visible only in the log plots, show the contribution 
from target windows. }
\label{fig:back}
\end{figure}

The physics asymmetry $\APHYS$ is formed from $\ACORR$,
\begin{equation}
\APHYS = \frac{K}{\pol}\frac{\ACORR - \pol\sum_{i} A_{i}f_{i}}{1-\sum_{i} f_{i}},
\end{equation}
with corrections for the
beam polarization $\pol$, background fractions $f_i$ with asymmetries $A_i$
and finite kinematic acceptance $K$. These corrections are
described below and summarized in
Table~\ref{table:Acorrections}. 
The first line lists the cumulative $\ABEAM$
corrections discussed above, scaled by $K/\pol$.

\begin{table}
\begin{tabular}{|l|rcl|rcl|}\hline
Correction (ppb) & \multicolumn{3}{c|}{Helium} & \multicolumn{3}{c|}{Hydrogen}\\ 
\hline \hline
Beam Asyms.    & $183$ & $\pm$ & $59$  & $-10$  & $\pm$ & $17$ \\
Target window bkg. & $113$   & $\pm$ & $32$  & $7$  & $\pm$ & $19$ \\
Helium QE  bkg.    & $12$  & $\pm$ & $20$ & & - & \\
Rescatter bkg.     & $20$  & $\pm$ & $15$  & $2$  & $\pm$ & \,\,\,$4$ \\
Nonlinearity     & $0$  & $\pm$ & $58$   & $0$     & $\pm$ & $15$\\
\hline \hline 
Scale Factor & \multicolumn{3}{c|}{Helium} & \multicolumn{3}{c|}{Hydrogen}\\ 
\hline \hline
Acceptance factor $K$     & $1.000$ & $\pm$ & $0.001$ & $0.979$ & $\pm$ & $0.002$ \\
$Q^{2}$ Scale      & $1.000$  & $\pm$ & $0.009$ & $1.000$  & $\pm$ & $0.017$ \\ 
Polarization $\pol$ & $0.844$ & $\pm$ & $0.008$ & $0.871$ & $\pm$ & $0.009$\\
\hline
\end{tabular}
\caption{Corrections to $\ACORR$ and systematic errors.}
\label{table:Acorrections}
\end{table}

A powerful feature of the apparatus is the spectrometers' ability
to focus the elastically scattered electrons into a compact region. Indeed,
much less than 1\%\ of the flux intercepted by the detectors originated 
from inelastic scattering in the target cryogen.
Figure~\ref{fig:back} shows charged particle spectra obtained 
with dedicated low-intensity runs and measured by drift chambers
in front of the detectors.
The dominant background was quasi-elastic
scattering from target windows, separately measured
using an equivalent aluminum target and computed to be $1.8 \pm 0.2$\%\ ($\he$) 
and $0.76 \pm 0.25$\%\ ($\h$).
 
An electron must give up more than 19~MeV to break
up the $^4$He nucleus and undergo quasi-elastic scattering off nucleons. 
Figure~\ref{fig:back} shows that the quasi-elastic threshold 
lies beyond the edge of the detector. A limit
of 0.15$\pm$0.15\% on this background was
placed by detailed studies of the low-intensity data. For $^1$H, the
$\pi^0$ threshold is beyond the extent of the plot; direct background 
from inelastic scattering is thus negligible.  

Background from rescattering in the spectrometer apertures was studied by 
varying
the spectrometer momentum in dedicated runs to measure inelastic spectra
and to obtain the detector response as a function of scattered electron energy
under running conditions. From these two distributions,
the rescattering background 
was estimated to be $0.25 \pm 0.15$\%\ ($\he$) 
and $0.10 \pm 0.05$\%\ ($\h$).  

For each source of background, a theoretical estimate for $\APV$ 
was used, with relative uncertainties taken to be $100\%$ or more 
to account for kinematic variations and resonance contributions. 
The resulting corrections and the associated errors are shown
in Table~\ref{table:Acorrections}. 
Upper limits on rescattering contributions from exposed 
iron in the spectrometer led to an additional uncertainty of $5$~ppb.

Nonlinearity in the PMT response was limited to 1\%\ in bench-tests that mimicked
running conditions. The
relative nonlinearity between the PMT response and those of the beam intensity
monitors was $<2$\%. 
A nuclear recoil technique using a water-cell target~\cite{happexhe}
was used to determine the scattering angle $\theta_{\rm{lab}}$,
thus keeping the scale error on  $\langle\qsq\rangle$ due to $\theta_{\rm{lab}}$
to be  $<0.2\%$.
The acceptance correction $K$ accounted for the 
non-linear dependence of the asymmetry with $\qsq$. 

The beam polarization, $\pol$, was continuously monitored by a Compton polarimeter;
results, averaged over the duration of each run, are
listed in Tab.~\ref{table:Acorrections}.
Redundant cross-calibration of the recoil Compton electron spectrum
restricted the relative systematic error to $\approx1$\%.
The results were consistent, within systematic uncertainties, with those 
obtained
from recoil Compton photon asymmetries, and with dedicated measurements
using M{\o}ller scattering in the experimental hall and Mott scattering 
at low energy. 
Throughout the asymmetry and background analysis, blinding
offsets were maintained on both results.  These offsets, which were significantly
larger than the respective statistical errors, were removed only after all 
analysis tasks were completed.
After all corrections: 
\begin{eqnarray*}
\APHYS^{\rm{He}} & = & +6.40 \pm 0.23 \,\,\mbox{(stat)} \,\pm 0.12 \,\,\mbox{(syst) ppm}, \\
\APHYS^{\rm{H}} & = & -1.58 \pm 0.12 \,\,\mbox{(stat)} \,\pm 0.04 \,\,\mbox{(syst) ppm}.
\end{eqnarray*}

The theoretical predictions $\ANS^{\rm{He}}$ and $\ANS^{\rm{H}}$ with $G^s=0$ 
were estimated using the formalism in~\cite{musolf94} and described 
in our previous publications~\cite{happexh,happexhe}. 
The electroweak radiative corrections, calculated using the $\MSbar$
renormalization scheme, introduced negligible uncertainties.

\begin{figure}
\includegraphics[width=1.0\columnwidth]{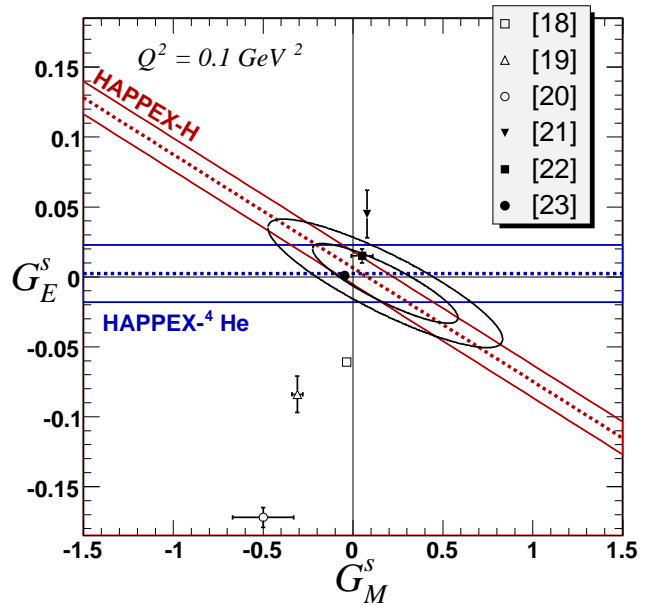}
\caption{68 and 95\%\ C.L. constraints in the $\ges-\gms$ plane from
data from this apparatus~(\cite{happexhe,happexh} and this Letter). 
Various theoretical predictions are plotted with published uncertainty
estimates, when available. 
The 1-$\sigma$ bands (a quadrature sum of statistical and systematic errors) 
and central values (dashed lines) from the new results alone are also shown.}
\label{fig:gegm}
\end{figure}

Assuming a pure isoscalar $0^+\rightarrow 0^+$ 
transition, $\ANS^{\rm{He}}$ is completely independent of 
nuclear structure and determined purely by electroweak parameters.
D-state and isospin 
admixtures and meson exchange currents are negligible at the level of the 
experimental fractional accuracy of $\sim 3$\%~\cite{donnellyetc}. For our
kinematics ($E_b$=2.75~GeV, $\langle\qsq\rangle=0.077\gevc$) we
obtain $\ANS^{\rm{He}}=+6.37$ ppm.

Electromagnetic form factors from a  phenomenological fit to the world 
data at low $\qsq$~\cite{friedrich} were used to calculate $\ANS^{\rm{H}}$, 
with uncertainties governed by data near \mbox{$\qsq\sim0.1\gevc$}.
The value used for $\gen=0.037$, with a 10\% relative 
uncertainty based on new data from the BLAST experiment~\cite{ziskin}.
For our kinematics ($E_b$=3.18~GeV, $\langle\qsq\rangle=0.109\gevc$) we
obtain $\ANS^{\rm{H}}=-1.66\pm0.05$ ppm.
This includes a contribution from the axial form factor $\gaz$, and associated
radiative corrections~\cite{zhu}, of $-0.037\pm 0.018$~ppm.

Comparing our results to the theoretical expectations, we extract 
$\ges= 0.002 \pm 0.014 \pm 0.007$ at $\qsq=0.077\gevc$ and 
$\ges+0.09\gms=0.007\pm 0.011 \pm 0.004 \pm 0.005$ (FF) at $\qsq=0.109\gevc$, 
where the uncertainties in the nucleon electromagnetic form factors govern
the last error.
Figure~\ref{fig:gegm} displays the combined result for these 
and our previous measurements~\cite{happexhe,happexh}, 
taken with $\langle\qsq\rangle$ between $0.077$-$0.109\gevc$.
The requisite small extrapolation to a common $\qsq=0.1\gevc$
was made assuming that $\ges\propto\qsq$ and that $\gms$ is constant.
The values $\ges=-0.005\pm 0.019$ and $\gms=0.18\pm 0.27$ 
(correlation coefficient =$-0.87$) are obtained.
The results are quite insensitive to variations in $\gaz$, as evidenced
by the negligible change induced by an alternate fit similar to
that in \cite{young}, where $\gaz$ is constrained
by other $\APV$ data.

Figure~\ref{fig:gegm} also displays predictions from selected theoretical 
models~\cite{park,drechsel,hammer,silva,lewis,leinweber}. Those 
that predict little strange quark dynamics in the vector form 
factors are favored~\cite{lewis,leinweber}. 
A global fit to all low-$\qsq$ measurements of $\ges$ and $\gms$,
similar to that performed in~\cite{young}, finds that 
other measurements~\cite{sample,A42,G0} which had suggested non-zero
strangeness effects are consistent, within quoted uncertainties,
with our results at $\qsq= 0.1\gevc$. 
Due to the improved statistical precision 
and lower $\gaz$ sensitivity of our result, adding these
other measurements in a global fit does not alter our conclusions.

In summary, we have reported the most precise constraints on the strange 
form factors at $\qsq\sim 0.1\gevc$. 
The results, consistent within errors 
with other $\APV$ measurements, leave little room for observable nucleon
strangeness dynamics at low $\qsq$. 
Theoretical uncertainties,
especially regarding 
the assumption of charge symmetry~\cite{CSV06}, preclude 
significant improvement to the measurements reported here.
While future experiments will pursue the search for non-zero strangeness
at higher $\qsq$, it now becomes a challenge for various theoretical
approaches to reconcile these results and enhance our understanding of 
nucleon structure.

\begin{acknowledgments}
We wish to thank the entire staff of JLab for their efforts to support this
experiment. 
This work was supported by The Southeastern Universities Research Association, Inc. 
under U.S. DOE Contract No. DE-AC05-84150, 
and by the DOE and NSF (United States), the INFN (Italy), and the CEA (France).

\end{acknowledgments}

\end{document}